\documentclass[journal=jacsat,manuscript=article]{achemso}

\usepackage[version=3]{mhchem} 

\usepackage{amsmath}
\usepackage{upgreek}
\usepackage{epstopdf}

\author{D. Schmidt}
\affiliation[TU Dortmund]{Experimentelle Physik 2, Technische Universit\"at Dortmund, D-44221 Dortmund, Germany}
\author{T. Godde}
\affiliation[University of Sheffield]{Department of Physics and Astronomy, University of Sheffield, Sheffield S3 7RH, UK}
\email{t.godde@sheffield.ac.uk}
\author{J. Schmutzler}
\affiliation[TU Dortmund]{Experimentelle Physik 2, Technische Universit\"at Dortmund, D-44221 Dortmund, Germany}
\author{M. A\ss{}mann}
\affiliation[TU Dortmund]{Experimentelle Physik 2, Technische Universit\"at Dortmund, D-44221 Dortmund, Germany}
\affiliation[University of Manchester]{School of Physics and Astronomy, University of Manchester, Manchester M13 9PL, UK}
\author{J. Debus}
\affiliation[TU Dortmund]{Experimentelle Physik 2, Technische Universit\"at Dortmund, D-44221 Dortmund, Germany}
\author{F. Withers}
\affiliation[University of Manchester]{School of Physics and Astronomy, University of Manchester, Manchester M13 9PL, UK}
\author{E. M. Alexeev}
\affiliation[University of Sheffield]{Department of Physics and Astronomy, University of Sheffield, Sheffield S3 7RH, UK}
\author{O. Del Pozo-Zamudio}
\affiliation[University of Sheffield]{Department of Physics and Astronomy, University of Sheffield, Sheffield S3 7RH, UK}
\author{O. V. Skrypka}
\affiliation[University of Sheffield]{Department of Physics and Astronomy, University of Sheffield, Sheffield S3 7RH, UK}
\author{K. S. Novoselov}
\affiliation[University of Manchester]{School of Physics and Astronomy, University of Manchester, Manchester M13 9PL, UK}
\author{M. Bayer}
\affiliation[TU Dortmund]{Experimentelle Physik 2, Technische Universit\"at Dortmund, D-44221 Dortmund, Germany}
\author{A. I. Tartakovskii}
\affiliation[University of Sheffield]{Department of Physics and Astronomy, University of Sheffield, Sheffield S3 7RH, UK}
\email{a.tartakovskii@sheffield.ac.uk}

\title{Exciton and trion dynamics in atomically thin MoSe$_2$ and WSe$_2$: effect of localization}

\begin{document}



\begin{abstract}
We present a detailed investigation of the exciton and trion dynamics in naturally doped MoSe$_2$ and WSe$_2$ single atomic layers as a function of temperature in the range 10-300K under above band-gap laser excitation. By combining time-integrated and time-resolved photoluminescence (PL) spectroscopy we show the importance of exciton and trion localization in both materials at low temperatures. We also reveal the transition to delocalized exciton complexes at higher temperatures where the exciton and trion thermal energy exceeds the typical localization energy. This is accompanied with strong changes in PL including suppression of the trion PL and decrease of the trion PL life-time, as well as significant changes for neutral excitons in the temperature dependence of the PL intensity and appearance of a pronounced slow PL decay component. In MoSe$_2$ and WSe$_2$ studied here, the temperatures where such strong changes occur are observed at around 100 and 200 K, respectively, in agreement with their inhomogeneous PL linewidth of 8 and 20 meV at $T\approx$10K. The observed behavior is a result of a complex interplay between influences of the specific energy ordering of bright and dark excitons in MoSe$_2$ and WSe$_2$, sample doping, trion and exciton localization and various temperature-dependent non-radiative processes. 
\end{abstract}



\section{Introduction}

Monolayers of semiconducting transition metal dichalcogenides (TMDCs) such as MoS$_2$, WS$_2$, MoSe$_2$ and WSe$_2$ have attracted considerable attention following the discovery of the indirect-to-direct bandgap transition \cite{MakPRL2010,SplendianiNanoLett2010,EdaNanoLett2011,WangNatNano2012} and the coupling of the spin and valley degrees of freedom in atomically thin layers \cite{XuNatPhys2014} and heterostructures \cite{RiveraScience2016}. In these materials, excitons exhibit very high binding energies of a few 100s of meV \cite{YeNature2014,HePRL2014,ChernikovPRL2014}, leading to their stability at room temperature attractive for optoelectronic devices \cite{PospischilNatNano2014,WithersNatMat2015,WithersNanoLett2015}. Furthermore, optical spectra of MoSe$_2$ and WSe$_2$, studied in this work, show well-resolved exciton and trion transitions \cite{RossNatComm2013,WangPRB2014,AroraNscale2015a,AroraNscale2015b,YouNatPhys2015,WithersNanoLett2015}, enabling new insights from the optics experiments into the physics of these materials.  

As demonstrated by theory \cite{XiaoPRL2012,Kormanyos2DMater2015,DeryPRB2015}, an important characteristic of these compounds is the strong spin-orbit interaction. It leads to the splitting between the dark and bright exciton sub-bands of the so-called A-exciton defined by the spin-orbit splitting in the conduction band,  $\Delta_{so}$. Here we refer to the excitons composed of the electron and hole with the same spin as 'bright', and with opposite spins as 'dark'. In both materials $|\Delta_{so}|$ of $\approx$30-40~meV is expected\cite{Kormanyos2DMater2015,DeryPRB2015}. However, $\Delta_{so}$ is predicted to be negative in WSe$_2$ leading to the lowest energy dark exciton sub-band in contrast to MoSe$_2$ where $\Delta_{so}$ is positive and the lowest exciton sub-band is bright\cite{Kormanyos2DMater2015,DeryPRB2015,ZhangPRL2015}. Such energy ordering of the bright and dark excitons have been shown to lead to the growth (decrease) of photoluminescence (PL) and electroluminescence (EL) intensity in WSe$_2$ (MoSe$_2$) with increasing temperature \cite{RossNatComm2013,WangPRB2014,AroraNscale2015a,AroraNscale2015b,YouNatPhys2015,WithersNanoLett2015}. 

In both MoSe$_2$ and WSe$_2$ the bright trion peak (labeled here as '$X^*$') is observed in absorption or PL about 30 meV below the bright exciton \cite{RossNatComm2013,WangPRB2014,AroraNscale2015a,AroraNscale2015b,YouNatPhys2015,WithersNanoLett2015,WangAPL2015,SinghPRB2016}. This is the lowest energy exciton complex in MoSe$_2$, whereas in WSe$_2$, the 'spin-forbidden' dark trion also exists at even lower energy. In this description we neglect the electron-hole exchange interaction, which may lead to a more complex energy spectrum, but is not expected to change considerably the ordering of the bands in these materials \cite{Kormanyos2DMater2015,DeryPRB2015,ZhangPRL2015}.

Owing to the particular ordering of the dark and bright exciton sub-bands, differences in the carrier dynamics in MoSe$_2$ and WSe$_2$ are expected, potentially impacting their spin and valley dynamics and performance in optoelectronic devices. Further dynamic effects are expected from trions\cite{SinghPRB2016} in our samples, where considerable electron densities are present and trion formation probability is therefore high. Finally, the impact of trion and exciton localization should be taken into account when considering low temperature behavior. So far, the interplay between these three factors, the specific band-structure, impact of trions and localization has not been considered in detail in recent time-resolved PL studies on monolayer TMDC semiconductors\cite{KornAPL2011,LagardePRL2014,YanAPL2014,ZhangPRL2015,AmaniScience2015,RobertPRB2016}.

In this work, we carry out a direct comparison between MoSe$_2$ and WSe$_2$ in samples with relatively high doping and clearly resolvable trion and exciton PL peaks, enabling different regimes of the temperature-dependent exciton and trion populations to be accessed. We identify two distinct regimes of high and low temperatures. In the former, in both MoSe$_2$ and WSe$_2$ the trion PL is suppressed, and the neutral exciton time-resolved PL (TRPL) shows bi-exponential decay in a wide range of temperatures with the slow component contribution increasing with $T$ and reaching nearly 100\% at room $T$, exhibiting the PL decay times of the order of 100 ps. We interpret the slow component as a sign of thermalization of excitons in the high-k reservoir states decoupled from light. In the low temperature regime, the trion PL intensity strongly increases with decreasing $T$ in both materials. PL dynamics shows several types of behaviour with typical life-times of several ps consistent with (i) exciton and trion energy relaxation into the lower energy states (trions and dark excitons, respectively); (ii) strong non-radiative decay for trions and excitons activated from the localizing disorder potential and thus having higher mobility and higher probability to reach non-radiative centres. An important observation enabling to link the observed behavior to the effect of localization is the difference in the 'threshold' temperatures where the behaviour changes from high to low temperature character: it is around 100 K for MoSe$_2$ and in excess of 200K for WSe$_2$. This correlates well with the observed inhomogeneous broadening of exciton and trion PL in MoSe$_2$ (8 meV) and WSe$_2$ (20 meV).

\section{Experimental procedure}

We investigate atomic monolayers of MoSe$_2$ and WSe$_2$ placed on a 10-20 nm layer of hBN. The whole structures are assembled on SiO$_2$ (290 nm) thermally grown on an n-doped Si substrate. The MoSe$_2$ samples were prepared with electrical contacts allowing electrostatic doping of the films by varying the gate voltage\cite{RossNatComm2013}. Further details on the sample fabrication can be found in Ref.~\cite{WithersNatMat2015,WithersNanoLett2015}.

The samples were mounted onto the cold-finger of a continuous flow cryostat, where temperature could be varied between 10 and 300 K.
A frequency-doubled Ti-Sapphire laser tuned to 3 eV was used for sample excitation with sub-ps pulses. The laser beam was focused under normal incidence onto the sample using a 50x microscope objective (NA 0.42), providing a 2 $\mu$m diameter excitation spot. PL from the sample was collected by the same objective. The time-integrated PL was measured using a spectrometer and nitrogen-cooled charged-coupled device, whereas for the time-resolved PL a streak camera was used. In the latter configuration, a liquid crystal filter (7~nm full width at half maximum) or a spectrometer equipped with 300 lines/mm grating was used to enable spectral resolution of the detected signal. The corresponding temporal resolution of 4~ps or 12~ps, respectively, was achieved. The set-up with 12 ps resolution was used to measure the PL decay curves in Fig.\ref{MoSe1}(d) only. The extracted life-times for these data are shown with open symbols in Fig.\ref{MoSe2}(b). Usually PL decay times, $\tau_{PL}$, shorter than the set-up resolution can be extracted by using convolution with the apparatus response function.

\section{Results}

\subsection{Temperature dependence of MoSe$_2$ photoluminescence}

\begin{figure}
\begin{center}
\includegraphics[width=\linewidth]{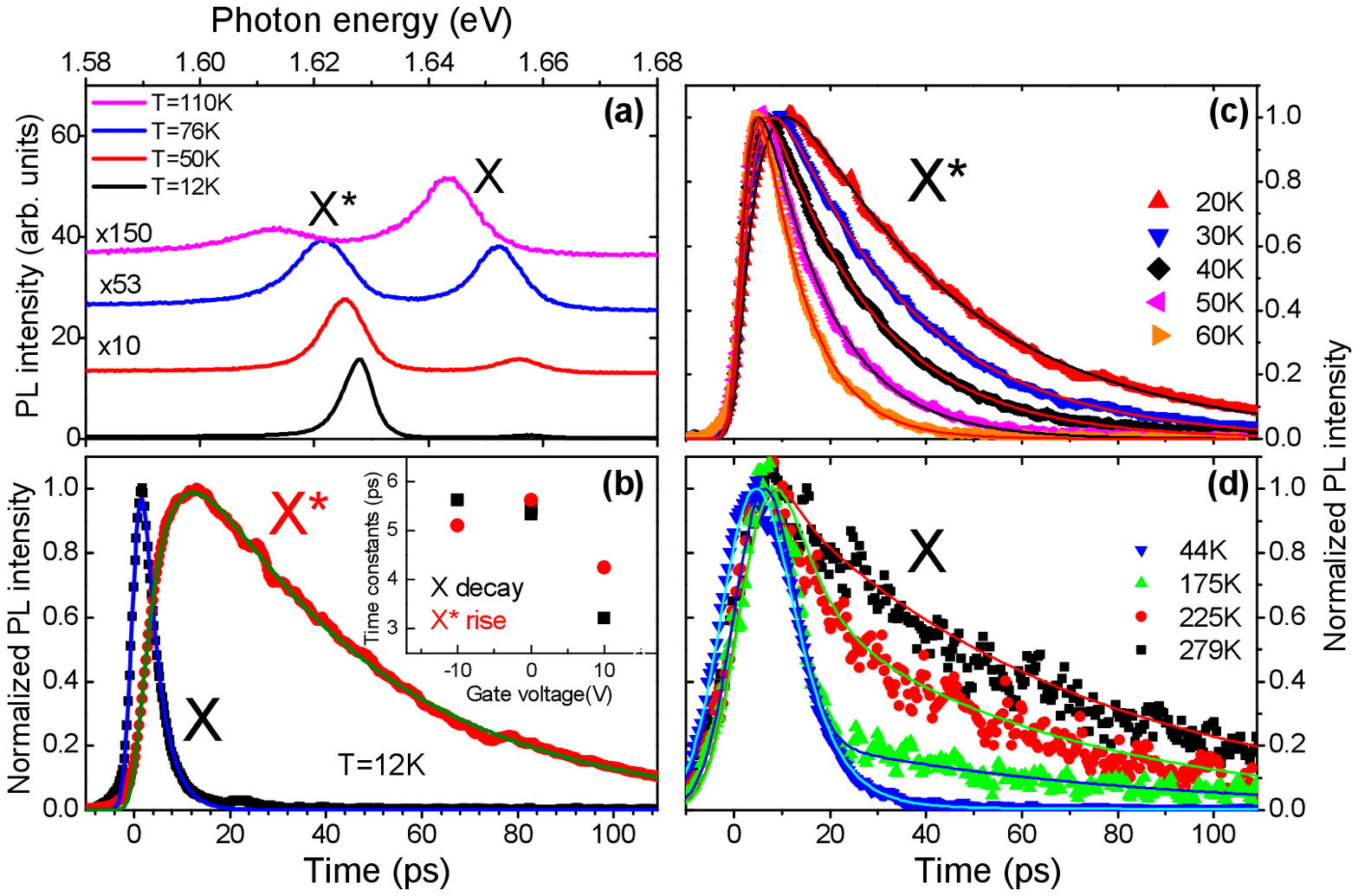}
\caption{Time-integrated (TI) and time-resolved (TR) photoluminescence (PL) for single-monolayer MoSe$_2$ samples. 
(a) TIPL spectra in the range 12-110K. Exciton ($X$) and trion ($X^*$) peaks are labeled. Spectra are multiplied by factors shown on the graph. 
(b) TRPL PL traces for $T$=12 K for $X$ (squares) and $X^*$ (circles) (sample 1). Lines show the fitting functions. The inset shows the $X$ PL decay time and $X^*$ rise time as a function of gate voltage in sample 5. The gate voltage is used to vary the carrier concentration in the film with a positive voltage corresponding to a higher concentration.
(c),(d) TRPL traces for $X^*$ (sample 1) in (c) and $X$ (sample 2) in (d) for a range of temperatures. Normalized PL intensities are shown.}
\label{MoSe1}
\end{center}
\end{figure}

In MoSe$_2$, carrier accumulation in the bright exciton states at low $T$ gives rise to strong PL as shown in Fig.~\ref{MoSe1}~(a). The two distinct peaks at 1.658 eV and 1.628 eV in the spectra measured with $P_{ex}$=7 $\mu$W for $T$=12 K correspond to the neutral exciton, $X$, and the trion, X$^*$. In the studied samples, the trion peak is more than one order of magnitude stronger than the neutral exciton peak. This reflects significant doping as well as efficient trion formation at low temperatures \cite{RossNatComm2013,AroraNscale2015b,WithersNanoLett2015}. 

\begin{figure}
\begin{center}
\includegraphics[width=0.5\linewidth]{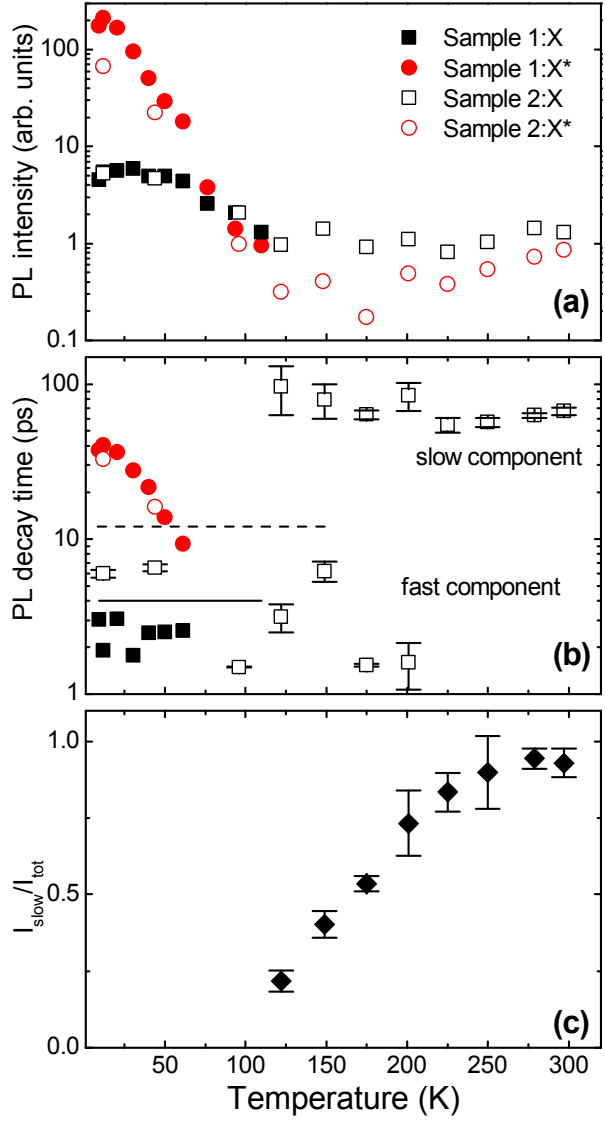}
\caption{Temperature dependence of time-integrated (TI) and time-resolved (TR) photoluminescence (PL) for single-monolayer MoSe$_2$ samples. 
(a) Spectrally integrated TIPL intensity. Dependences for $X$ (squares), $X^*$ (circles) are shown for samples 1 and 2. 
(b) $X$ (squares) and $X^*$ (circles) PL decay times in samples 1 (solid symbols) and 2 (open symbols). The horizontal lines mark the set-up resolution for the two experiments: the dashed line is for the 12 ps resolution corresponding to the data shown with the open symbols; the solid line is for 4 ps resolution corresponding to the data shown with the solid symbols.
(c) The ratio of the time-integral of the function used to fit the slow PL decay component, $I_{slow}$, and the total integral under the PL decay curve, $I_{tot}$,    $r=I_{slow}/I_{tot}$.}
\label{MoSe2}
\end{center}
\end{figure}

With increasing temperature, the exciton PL increases relative to the trion PL [Fig.~\ref{MoSe1}~(a)], with the latter showing a dramatic reduction by up to 100 times in the range from 10 to 100~K, as further shown for two different samples in Fig.~\ref{MoSe2}~(a). This reduction is rather surprising given that the trion binding energy is around 30 meV, as observed from PL spectra in Fig.~\ref{MoSe1}~(a)\cite{RossNatComm2013,AroraNscale2015b,WithersNanoLett2015}, and trion dissociation should still be relatively ineffective for $T\approx$100~K. As seen in Fig.~\ref{MoSe1}~(a) and \ref{MoSe2}~(a), the $X$ PL shows a moderate decrease by a factor of 4 in the range 10K$<T<$100K, and then remains almost unchanged up to 300 K. At $T>100$K, the exciton peak has a low energy tail that can in principle be interpreted as weak $X^*$ PL. The integrated intensity of this shoulder is shown with red open symbols in Fig.~\ref{MoSe2}~(a). This interpretation and accuracy of the fit becomes less reliable at higher $T$, where $X$ has a relatively large width of $\approx$40 meV. 

To gain insights into the carrier dynamics, we performed TRPL measurements at various temperatures. Shown in Fig.~\ref{MoSe1}~(b) are typical PL transients of the $X$ and X$^*$ at $T$=12 K. Here, a sequential build-up of the transients is observed, i.e. the maximum of the trion PL is reached when the exciton PL has completely decayed. While the exciton PL dynamics are on the timescale of the temporal resolution of our setup with no resolvable rise, the trion dynamics comprise a resolvable rise and a decay. This indicates that the low energy neutral exciton states are populated extremely fast by the non-resonant optical excitation. The interaction with resident electrons of this exciton reservoir then leads to creation of trions. Gate-voltage-dependent TRPL measurements on a sample with electrical contacts (sample 5) show an increase of the $X$ PL decay time for decreasing carrier concentrations, as depicted in the inset in Fig.~\ref{MoSe1}~(b). As seen in this figure, the $X^*$ PL rise time follows a similar trend. This implies that the trion formation may be one of the dominant neutral exciton decay channels at low $T$. Recent pump-probe studies\cite{SinghPRB2016} found that at low temperatures the trion formation time $\approx$2~ps, which is in a reasonable agreement with the trion rise time observed here. 

Fig.~\ref{MoSe1}~(c) shows $X^*$ PL decay curves measured as a function of temperature for $T\leq$60K, in the range where $X^*$ PL can be measured reliably. The curves show a constant trend of the decay time shortening with $T$. Fig.~\ref{MoSe1}~(c) also shows shortening of the PL rise time, which may be a consequence of a faster decay. The decay can be described with a single-exponent function. The analysis of these data presented in Fig.~\ref{MoSe2}~(b) shows that the $X^*$ decay time decreases strongly from 41 ps at $T$=12 K to 9 ps at $T$=61 K. This evidence combined with the significant reduction in the time-integrated PL with $T$ indicates that in this temperature range trions experience highly efficient non-raditive decay.

A different behaviour is observed for the $X$ PL dynamics in Fig.~\ref{MoSe1}~(d), where the PL decay curves measured in a wider range of temperatures 44$\leq T\leq$279K are presented. Fast single exponential decay (below our resolution) is observed at low $T<$100K. The exciton dynamics becomes more complex for $T>$100K. Here, occurrence of a new slow PL decay component is observed. The corresponding decay times are presented in Fig.~\ref{MoSe2}~(b), where the horizontal lines mark our set-up resolution for the two experiments: the dashed line is for the 12 ps resolution corresponding to the data shown with the open symbols; the solid line is for 4 ps resolution corresponding to the data shown with the solid symbols. By comparing the two sets of data, it is seen that the accuracy in determination of the lifetime for the fast PL decay component is limited due to relatively poor time-resolution, which is also likely to be the source of the scatter in the data obtained using convolution with the response function of the apparatus. On the other hand, the measurement of the slow decay component is significantly less affected by the resolution. The corresponding decay time shows no apparent dependence on $T$, and is in the range 70-100 ps, reaching 70 ps at room temperature. It is evident from the PL transients in Fig.~\ref{MoSe1}(d), that the slow decay component becomes more pronounced as $T$ is increased. We quantify this with the ratio $r=I_{slow}/I_{tot}$. Here $I_{tot}$ is the total integral under the PL decay curve, and $I_{slow}$ is the time-integral of the fitting function used for the slow component. The temperature dependence of $r$ is plotted in Fig.~\ref{MoSe2}(c). A steep transition to the regime where the slow component dominates occurs between 100 and 250 K (e.g. $r>0.8$ for $T$=225K), in the regime where the trion contribution to PL becomes negligible.

\subsection{Temperature dependence of WSe$_2$ photoluminescence}

\begin{figure}
\begin{center}
\includegraphics[width=\linewidth]{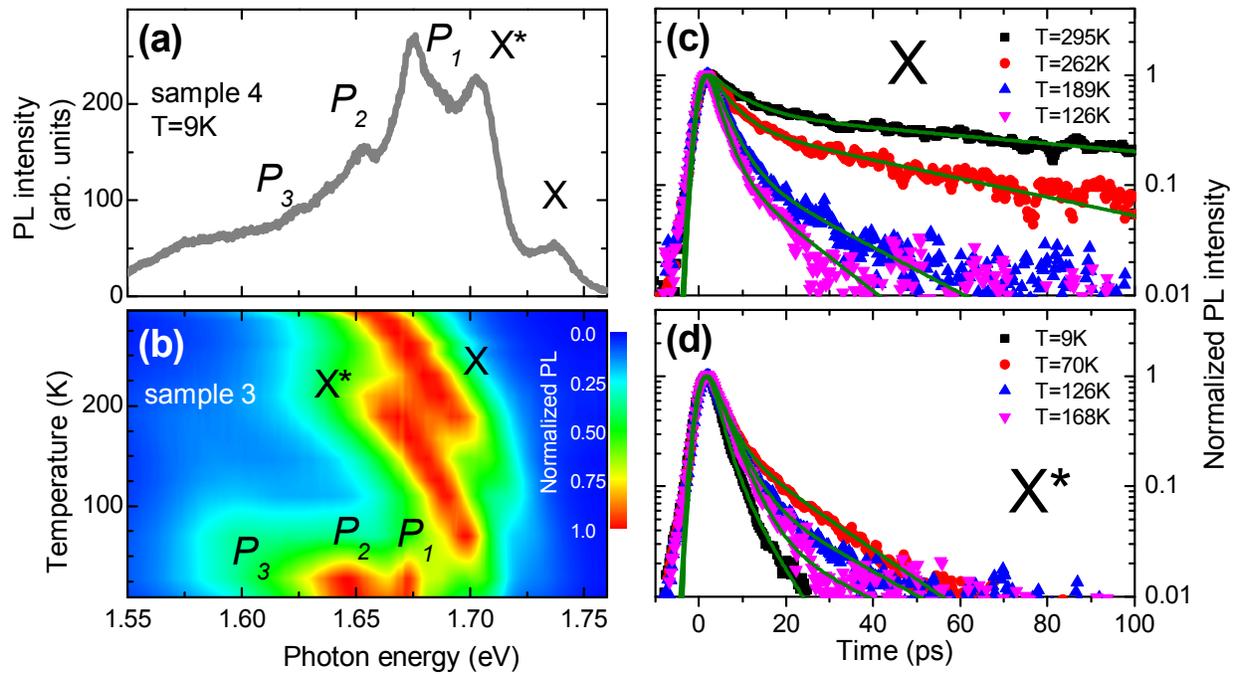}
\caption{Time-integrated (TI) and time-resolved (TR) photoluminescence (PL) for single-monolayer WSe$_2$ samples. 
(a) A TIPL spectrum measured at $T$=9K in sample 4. Neutral exciton ($X$) and trion ($X^*$) peaks as well as localized exciton peaks P$_1$, P$_2$ and P$_3$ are labeled. 
(b) TIPL in a temperature range 9-293~K. Individual spectra at each temperature are normalized. Features as in graph (a) are labeled. 
(c),(d) TRPL traces for $X$ (c) and $X^*$ (d) in sample 3 for a range of temperatures. Normalized PL intensities are shown.}
\label{WSe1}
\end{center}
\end{figure}

\begin{figure}
\begin{center}
\includegraphics[width=0.5\linewidth]{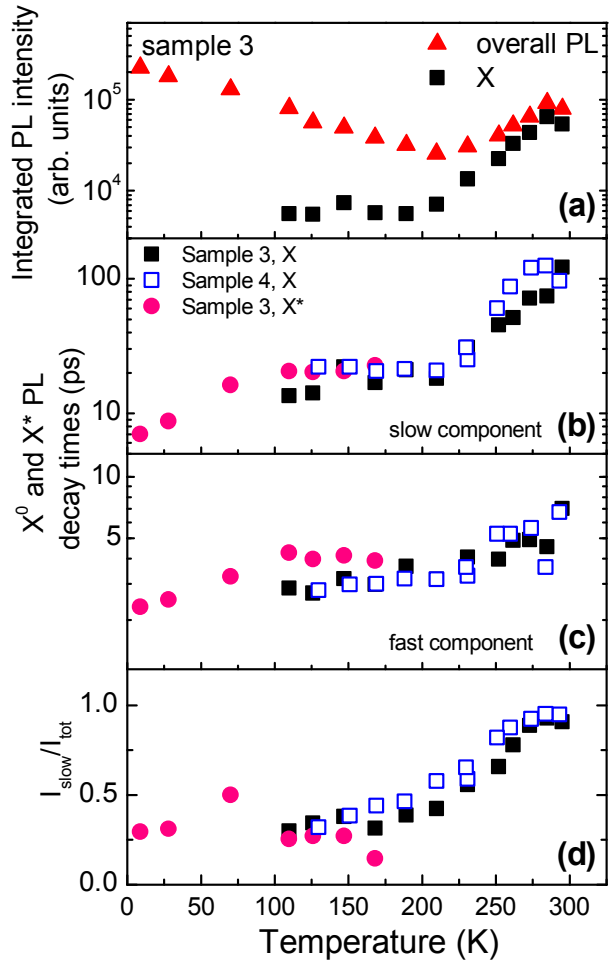}
\caption{Temperature dependence of time-integrated (TI) and time-resolved (TR) photoluminescence (PL) for single-monolayer WSe$_2$ samples. 
(a) Spectrally integrated TIPL intensity. Dependences for $X$ (squares) and all PL in the range shown in Fig.\ref{WSe1} (triangles) are shown. 
(b),(c) $X$ (squares) and $X^*$ (circles) PL decay times. The data for the slow and fast decay components are shown in (b) and (c), respectively. Open and solid symbols show $X$ PL decay data for two different samples.
(d) The ratio of the time-integral of the function used to fit the slow PL decay component, $I_{slow}$, and the total integral under the PL decay curve, $I_{tot}$,    $r=I_{slow}/I_{tot}$. Data for both $X$ (squares) and $X^*$ (circles) are shown.}
\label{WSe2}
\end{center}
\end{figure}

Similarly to single-layer MoSe$_2$, WSe$_2$ exhibits readily identifiable PL peaks at low $T$. Fig.~\ref{WSe1}~(a) shows a typical low $T$=9~K PL spectrum (measured on sample 4) with four pronounced maxima at 1.737 eV ($X$), 1.704 eV ($X^*$), 1.675 eV ($P_1$) and 1.653 eV ($P_2$), as well as a broad feature at lower energy ($P_3$). The two highest energy peaks are attributed to the neutral exciton and the corresponding trion \cite{WangPRB2014,AroraNscale2015a,YouNatPhys2015}. The lower energy peaks are typically assigned to bound excitons \cite{WangPRB2014,AroraNscale2015a,YouNatPhys2015,WithersNanoLett2015}. Their PL decreases rapidly with increasing temperature, and is negligible for $T >$100 K \cite{WangPRB2014,AroraNscale2015a,YouNatPhys2015,WithersNanoLett2015}, as shown in Fig.~\ref{WSe1}~(b) (data for sample 3). In our sample, possibly due to significant doping, $X^*$ dominates the PL spectrum between 70 and 200K. However, its intensity steadily decreases with $T$ above 70K. As seen in Fig.~\ref{WSe1}~(b), for $T >$230 K the neutral exciton PL dominates. Fig.~\ref{WSe2}~(a) also shows that the PL intensity from all states (referred to as 'overall' in the figure) decreases with $T$ up to about 200K. However, the strong increase of the neutral exciton PL for $T>200$K up to ambient conditions overrides this trend. This behavior strongly contrasts WSe$_2$ with MoSe$_2$. 

Because of the dominating $X^*$ PL at low $T$ and relatively broad PL features, the TRPL signal for the neutral exciton can be measured reliably only for temperatures above 110 K [see Fig.~\ref{WSe1}~(c)]. A bi-exponential decay is observed with clearly distinct 'fast' and 'slow' components. The contribution of the slow component increases with $T$, accompanied with the increase of the associated PL decay time. $X^*$ also shows bi-exponential decay in the whole range of $T$, where it is detectable for $T<$150K [see Fig.~\ref{WSe1}~(d)]. The slow component is less pronounced than for $X$ at high $T$. 

Fig.~\ref{WSe2}(b),(c) show the analysis of these data. The trends for decay times in both fast and slow components are rather similar, with both gradually increasing with $T$. For $X^*$ (circles in Fig.~\ref{WSe2}), the fast decay times remain around the resolution limit with a slight increase up to 4 ps at $T>$100K, whereas the slow decay time grows from 7 ps for $T$=9 K to 23 ps for $T$=168K. The decay times for the slow and fast components for both exciton and trion are very similar in the overlapping region of 100K$>T>$175K, where both PL peaks are clearly detectable. For $X^*$, TRPL can be measured reliably up to approximately 170K, where contribution from the strong $X$ PL becomes difficult to separate. 

For higher temperatures 200K$>T>$300K, exciton decay times continue to grow: very slight increase from 3.6 to up to 5-6 ps is observed for the fast component, whereas a much sharper increase from 20 to 120 ps is observed for the slow component. At the same time, the contribution of the slow component to the overall exciton PL decay also increases. Similarly to the case of MoSe$_2$, we quantify the increase with $T$ of the contribution of the slow-decaying component by plotting in Fig.~\ref{WSe2}~(d) the ratio $r$. This shows that the contribution of the slow component dominates the radiative decay for $T>$200 K. From comparison with Fig.~\ref{WSe2}(a), it is evident that the range of temperatures where the rise of the exciton decay time is observed coincides with the temperature range where also both the slow decay component contribution and exciton PL is strongly increasing, whereas the trion PL becomes negligible. In contrast to $X$, it is evident from Fig.~\ref{WSe2}~(d) that the fast decay dominates the $X^*$ dynamics for all $T$.

\section{Discussion}

In order to interpret our data we consider the impact of exciton and trion localization on the PL intensities and dynamics. For $T\approx$10K, the typical PL full width at half maximum in the samples studied here are 8 meV ($X$) and 6 meV ($X^*$) for MoSe$_2$ and $\approx$20 meV (for both $X$ and $X^*$) in WSe$_2$. These values corresponding to the inhomogeneous broadening provide an approximate measure of the localization potential induced by the disorder. It could be expected that a 'transition' from localized to free excitons and trions may occur at temperatures of the order of 80 and 200 K in MoSe$_2$ and WSe$_2$, respectively. This is close to the temperatures where significant changes are observed in the temperature dependences of exciton and trion PL intensities and dynamics in MoSe$_2$ and WSe$_2$ in this study. 

In the low temperature regime probed in our experiments at $T\approx$10K, both neutral excitons and trions could form a mixture of localized and free complexes \cite{HegartyPRB1984}, with trions more likely to be localized due to a larger mass and their charge. In MoSe$_2$, the fast decay exhibited by excitons is caused by radiative recombination \cite{RobertPRB2016} and binding into trions as Fig.\ref{MoSe1}(b) implies. The PL decay is likely to reflect the dynamics of free excitons, as they have a higher oscillator strength compared to the localized states. Trion PL comes mostly from localized complexes and shows a longer decay time as a result. In this temperature regime in WSe$_2$, additional decay of trions to low energy dark states occurs, which may account for the relatively fast PL decay compared to MoSe$_2$.

When the temperature is increased, the fractions of both free trions and free excitons increase. The excitons and trions become more mobile which causes a new non-radiative decay as they can more easily reach non-radiative centers. This effect is very pronounced in MoSe$_2$ in the range $10<$T$<100$K where both trion and exciton PL intensity falls with $T$ with the overall PL intensity decrease by more than 100 times. In addition to weaker localization, the trion formation from free excitons and electrons is less likely as the temperature is increased\cite{RossNatComm2013}. Overall, in MoSe$_2$ the trion life-time shortens in this regime due to activation of localized states, ionization of the trions and occurrence of new non-radiative recombination channels. 

In WSe$_2$, the trion PL decrease gradually happens over a larger temperature range up to $\approx$250K, consistent with a larger thermal energy required to overcome the localization potential in this material. The trion PL decay shows a dominating and relatively weakly temperature-dependent fast component with the PL decay time of $\approx$5ps, possibly due to fast relaxation into the low energy dark states. In addition, there is a slow PL decay component with the PL decay time reaching up to 20 ps for $T>$100K, which may reflect decay of the neutral exciton population. The decay in those states could be observable in the trion PL decay due to their feeding into the trion population as well as a partial spectral overlap of excitons and trions. 

In the high $T$ regime, localization is not important and the probability to form trions is also low. Both factors lead to suppression of trion PL. In WSe$_2$ this is also accompanied with a significant growth with $T$ of the exciton PL intensity. This occurs due to thermalization leading to enhanced population of the bright exciton states, and also implies that in the presence of trions a strong non-radiative decay channel existed through which the exciton population could decay. Further to the PL intensity changes, the exciton PL dynamics exhibit a slow component caused by the exciton thermalization within the bright sub-band. This is accompanied with a spread of excitons over a range of k-vectors where they are decoupled from light. This could lead to a linear increase of the average exciton radiative life-time with $T$ \cite{RobertPRB2016}. In our experiments, in MoSe$_2$ the decay time associated with the slow component is nearly constant over a wide range of temperatures (shows a weak decrease with $T$), which is possibly an indication that another temperature-dependent non-radiative decay mechanism is present influencing the neutral exciton population. In WSe$_2$, a strong increase of PL decay time is observed once the exciton thermal energy has exceeded the typical localization energy of 20 meV for $T>$200K. However, in neither MoSe$_2$ or WSe$_2$, the exciton decay time reaches the ns range recently reported in other studies at elevated temperatures\cite{AmaniScience2015}. The most likely cause of such relatively short life-times is the exciton-exciton annihilation \cite{KumarPRB2014,MouriPRB2014,SunNanoLett2014,AmaniScience2015}, more pronounced for free excitons and elevated temperatures.

\section{Conclusions}

In conclusion, we presented a detailed study of the carrier dynamics in atomically thin semiconductors MoSe$_2$ and WSe$_2$ in a wide temperature range 10-300 K, revealing a complex interplay between the particular ordering of the dark and bright exciton sub-bands in the two materials as well as presence of trions, influence of exciton and trion localization and various non-radiative processes. Despite significant differences in the band-structure leading to the opposing temperature dependences of the neutral exciton PL intensity in MoSe$_2$ and WSe$_2$, many similarities have been observed in the neutral exciton dynamics such as a pronounced increase of the exciton decay time at high temperatures. It is observed that major changes in the neutral exciton dynamics, as well as exciton and trion PL intensity, occur for elevated temperatures where the exciton thermal energy exceeds the typical localization energy caused by the disorder potential. This transition to the mobile excitons occurs in our samples around 100 and 200 K in MoSe$_2$ and WSe$_2$, respectively, in agreement with the differing inhomogeneous linewidths observed in PL. Complex trion and exciton dynamics are observed for lower temperatures where thermal energy is of the order or lower than the typical localization energy. 

The impact of localization requires further study, as it fundamentally defines the character of electronic excitations in the two-dimensional system, which in addition to trions in the low electron density and/or strong localization limit can also exhibit Fermi polarons \cite{SidlerArxiv2016} in the opposite limit. In the context of TMDCs, the ability to uncover this character will help to reveal the origin of the unexpected values for exciton and trion g-factors observed recently \cite{LiPRL2014,AivazianNatPhys2015,SrivastavaNatPhys2015,Wang2DMater2015,SmolenskiPRX2016}, opening the way to implementation of valleytronics ideas \cite{XuNatPhys2014}.

\begin{acknowledgement}

We thank the financial support of the Graphene Flagship, FP7 ITN S$^3$NANO, the EPSRC Programme Grant EP/J007544/1 and grant EP/M012727/1. FW acknowledge support from Royal Academy of Engineering. KSN also acknowledges support from ERC, EPSRC (Towards Engineering Grand Challenges), the Royal Society, US Army Research Office, US Navy Research Office, US Airforce Research Office. \\ 

\end{acknowledgement}

%

\providecommand{\latin}[1]{#1}
\providecommand*\mcitethebibliography{\thebibliography}
\csname @ifundefined\endcsname{endmcitethebibliography}
  {\let\endmcitethebibliography\endthebibliography}{}

\newpage

\newpage

\newpage

\newpage

\end{document}